# Critical length limiting super-low friction

Ming Ma[1], Andrea Benassi[2], Andrea Vanossi[3,4], Michael Urbakh[1,*]

[1] School of Chemistry, Tel Aviv University, 69978 Tel Aviv, Israel

[2] Empa, Swiss Federal Laboratories for Materials Science and Technology, CH-8600 Dübendorf, Switzerland.

[3] CNR-IOM Democritos National Simulation Center, Via Bonomea 265, 34136 Trieste, Italy

[4] International School for Advanced Studies (SISSA), Via Bonomea 265, 34136 Trieste, Italy

[*] Correspondence should be sent to `urbakh@post.tau.ac.il`

**Abstract**

Since the demonstration of super-low friction (superlubricity) in graphite at nanoscale, one of the main challenges in the field of nano- and micro-mechanics was to scale this phenomenon up. A key question to be addressed is to what extent superlubricity could persist, and what mechanisms could lead to its failure. Here, using an edge-driven Frenkel-Kontorova model, we establish a connection between the critical length above which superlubricity disappears and both intrinsic material properties and experimental parameters. A striking boost in dissipated energy with chain length emerges abruptly due to a high-friction stick-slip mechanism caused by deformation of the slider leading to a local commensuration with the substrate lattice. We derived a parameter-free analytical model for the critical length that is in excellent agreement with our numerical simulations. Our results provide a new perspective on friction and nano-manipulation and can serve as a theoretical basis for designing nano-devices with super-low friction, such as carbon nanotubes.

One of the most intriguing concepts of modern tribology, with exciting theoretical issues and possible practical applications, is the idea of interface frictionless sliding in incommensurate contact, named superlubricity [1-9]. In such a geometrical configuration, the lattice mismatch (or misalignment) can prevent interlocking and collective stick–slip motion of surface atoms, with a consequent vanishingly small frictional force, provided stiff enough substrates compared to their mutual interaction. The search and design of superlubric interfaces is a subject of practical importance in nano- and micro-mechanics, aimed to significant reduction of friction and wear in mechanical devices functioning at various scales.

Until very recently, superlubricity has been observed only on the nanoscale [1, 7]. Considerable progress has been achieved in 2012 when superlubricity was found for graphite samples of micrometer size [10]. The main challenge, however, is to scale this process up. With the capability of synthesizing and manipulating quasi 1D atomically perfect objects of large length, such as telescopic nanotubes [11], graphene nanoribbons [12] or aromatic polymers [13], nanotechnologies open nowadays the possibility to transpose the peculiar nanoscale tribological properties, such as superlubricity, to larger scales and exploit them to control sliding friction. Recently, a breakthrough has been achieved, demonstrating existence of superlubricity for centimeter-long double walled carbon nanotubes (DWCNTs) [11].

The robustness of the superlubricity phenomenon, however, remains a challenge. Even in clean wearless friction experiments with perfect atomic structures, superlubricity at large scales may be broken due to elasticity of contacting samples [14-16]. Thus, a key question to be addressed is to what extent superlubricity could persist, and what sort of mechanisms could lead to its failure. Previous scaling analysis of structural lubricity [14] was based on the assumption that a slider of length $L$ can be considered as essentially rigid and exhibits super-low kinetic friction as long as its stiffness on that length scale is larger than the corresponding interfacial stiffness. Here, we demonstrate that this assumption does not apply for the edge-driven configuration where the pulling or pushing force is applied to the edge of the slider. This configuration is typical for many frictional and nano-manipulation experiments performed with atomic force microscope, as sketched in Fig.1(a).

In this Letter, in the framework of an edge-driven Frenkel-Kontorova modeling approach [17, 18], we establish a clear connection between the critical length above which the super-low frictional regime disappears and the intrinsic material properties of the system. The transition to high-dissipative stick-slip regime is caused by the increase of the friction force with the chain length and the occurrence of an abrupt and significant slider deformation leading to a local commensuration with the underneath substrate lattice.

We model the quasi 1D systems depicted in Fig.1 with a chain of $N$ particles of mass $m$, linked by springs $K$, having rest length $a_c$, and driven on a periodic potential with periodicity $a_s$ and amplitude $U_0$. As shown in Fig. 1(b), the chain rightmost particle, with coordinate $X_N(t)$, is pulled at constant velocity $V_0$ through a spring $K_{dr}$ representing the lateral stiffness of the cantilever. The friction force is measured through the spring elongation, as $F = K_{dr}(V_0 t - X_N)$. If we express particle coordinates in units of $a_c$, $x_i \equiv X_i/a_c$, time in units of $\tau_0 \equiv \sqrt{m/K}$ and energies in units of $K a_c^2$, the dimensionless equations of motion become:

$$\begin{cases} \ddot{x}_i = F_i^{sub} + F_i^{el} - \gamma \dot{x}_i & for\ i < N \\ \ddot{x}_N = F_N^{sub} + F_N^{el} - \gamma \dot{x}_N + k_{dr}(v_0 \tau - x_N) & for\ i = N \end{cases} \quad (1)$$

where $F_i^{sub} = -u_0 \frac{2\pi}{(a_s/a_c)} \cos\left(\frac{2\pi x_i}{(a_s/a_c)}\right)$ is the force describing interaction between the $i$-th particle and substrate, $F_i^{el} = (x_{i+1} + x_{i-1} - 2x_i)$ for $1<i<N$, $F_1^{el} = (x_2 - x_1 - 1)$ and $F_N^{el} = -(x_N - x_{N-1} - 1)$ are elastic forces between the particles in the chain, and $u_0 = U_0/(K a_c^2)$, $v_0 = V_0(\tau_0/a_c)$, $k_{dr} = K_{dr}/K$. The viscous damping term, $\gamma = \eta \tau_0$, with $\eta$ being the dimensional damping coefficient, accounts for the driven particle energy dissipated into the substrate microscopic degrees of freedom.

In this work, we focus on stiff incommensurate systems with an interparticle stiffness $K$ sufficiently larger than the atomic-scale interfacial interaction, $K_{int} = U_0(2\pi/a_s)^2$; a physical condition met by many materials, including the quasi 1D structures depicted in Fig.1a. This defines, in the thermodynamic limit ($N \to \infty$), a continuum set of ground states that can be reached adiabatically through nonrigid displacement of chain atoms at no energy cost with a consequent vanishing static friction [2, 3].

The simulations have been performed for the incommensurate ratio $(a_s/a_c) = \frac{1+\sqrt{5}}{2}$, corresponding to the golden mean, and for a broad range of other parameters ($2\times10^{-3}<u_0<5\times10^{-2}$, $1.0\times10^{-3}<v_0<5.0\times10^{-2}$, $3.0\times10^{-3}<\gamma<7.0\times10^{-2}$ and $1\times10^{-2}<k_{dr}<0.2$). Similar results, as shown in Supplementary materials [19], have been found for inverse ratio of $a_s$ and $a_c$, corresponding to a 'rarefied' anti-soliton chain structure. The golden mean represents the 'most irrational' incommensurability, and thus, the most superlubric configuration [20]. The equations of motion have been integrated numerically using a Velocity-Verlet algorithm, focusing on the case at $T$=0. We have also performed simulations for finite temperature using a Langevin dynamics simulation, finding, however, that up to $T$=0.4$U_0/k_B$, the thermal effects do not lead to qualitative changes. For DWCNTs the ambient temperature is included in this range of parameters.

In Fig. 2 we present examples of frictional response for chains of different lengths. Here we show results for chains long enough ($N \gtrsim 200$, for our choice of parameters) to neglect any appreciable boundary effects. Figure 2(a) presents $F$ as a function of time for incommensurate chains of two different lengths. Waiting long enough from the initiation of sliding, a steady state is reached by both chains, however the final value of the friction force is significantly different. Moreover, while for the short chain the transition to the steady state is smooth, for the long one the friction force experiences a sudden jump almost doubling its value.

The average friction force per particle $\langle F \rangle/N$, calculated once the steady sliding state is reached, is plotted in Fig. 2(b) as a function of the chain length. It exhibits a sudden jump around $N_{cr} \approx 800$ increasing by almost a factor of two. For shorter chains, $200 \lesssim N < N_{cr}$, we found a super-low friction state, where the friction force is only slightly higher than the total "viscous" contribution to friction $N\eta V_0$. The latter gives the lower bound of the dynamic friction force expected for an ideal incommensurate contact. In the interval $200 \lesssim N < N_{cr}$, the force $\langle F \rangle / N$ only slightly changes with $N$. For chains longer than $N_{cr}$ a new channel of dissipation sets up, and superlubricity disappears. The transition from super-low to high friction state, as the chain length grows, is clearly related to the onset of the sudden jumps in the instantaneous force shown in Fig. 2(a) and triggered by a mechanical instability occurring for the edge-driven configuration as $N$ exceeds $N_{cr}$.

To give a quantitative measure of the degree of commensurability of the particles with respect to the substrate potential, we introduce the normalized distances between particles, $d(i) = (X_i - X_{i-1})/a_S$, where $i = 1,\ldots, N$ is the particle index. First, we focus on the dynamics at the $K_{dr}$-driven edge of the chain. Figure 3(a) presents the time dependence of the distance between the two rightmost particles, $d(N)$, for the chain of length $N = 1{,}000$ that shows the sudden jump in the instantaneous friction force, the latter is also plotted as a reference. The correlation between the two curves is evident, indicating that the friction increase is related to the nucleation of a commensurate region at the leading edge of the chain. A detailed picture of the ongoing phenomena can be obtained looking at the 2D maps for distances between particles, $d(i)$, along the chain as functions of the particle index and time, which are shown in Figs. 3(b),(c). For $N = 600$ during the frictional motion the whole chain remains in the incommensurate state with all inter-particle distances $d(i) < 1$. In the steady state regime of motion the distances $d(i)$ increase linearly, on average, with the particle index from the almost unstretched length at the trailing edge to the strongly stretched length at the leading edge. In contrast, for $N = 1{,}000$ a narrow region of particles forming the commensurate structure nucleates at the leading edge and propagates inside the chain in the direction opposite to the sliding motion (movie in [19]). Upon reaching the steady sliding, the commensurate region stabilizes at the critical length, $N_{cr}$, from the trailing edge forming a sharp boundary (domain wall) between the phases with $d(i) < 1$ and $d(i) > 1$. Panel 3(c) inset shows the interparticle distance distribution versus the particle index calculated in the steady regime for the chains of $N = 600$ and $1,000$. For $N=1,000$, as for shorter chains, the length, $d(i)$, grows linearly with $i$ for $i < N_{cr}$, however at $i = N_{cr}$ an abrupt jump in $d(i)$ occurs manifesting the formation of the domain wall. After the jump, $d(i)$ continues to grow linearly with $i$, keeping the same slope in both regions $i < N_{cr}$ and $i > N_{cr}$.

The particles located in the narrow commensurate region perform highly dissipative stick-slip motion, thereby greatly increasing the average friction force of the chain. This is highlighted in Fig. 3(d) showing the normalized average dissipated power, $P(i) = m\eta \lim_{T\to\infty} \frac{1}{T}\int_{t_0}^{T}(\dot{X}_\iota)^2 dt/(m\eta V_0^2)$, along the chain in the steady sliding regime. Here $t_0$ is the time corresponding to the onset of the steady state regime. $P(i)$ exhibits a high peak localized in the commensurate region of the chain, where it is two orders of magnitude higher than in the rest of the chain. This peak localized at the domain wall is characteristic for the mechanism of transition from the super-low to high friction discussed here, and it is absent for chains shorter than $N_{cr}$. In the range of parameters studied here, the peak value scales linearly with the strength of the particle-substrate interaction, $U_0$, and it is independent of the stiffness of the chain, $K$. In contrast, The width of the peak increases with $K$ and only slightly depends on $U_0$. In agreement with experimental observations for multi-walled CNTs [21], a slightly larger dissipation, compared to the rest of the chain, has been also found at the edges of the chain. This effect is independent of the chain length and results from 'translational symmetry' breaking at the edges.

Our simulations demonstrate that the abrupt jumps in the friction force occur also at the transitions corresponding to the nucleation of the commensurate states with distances between the chain particles equal to $2a_s$, $3a_s$ ... . However, these transitions require very large stretching of the chain that is beyond the linear elastic description.

In the steady state regime of motion the time-averaged friction force can be written as [22]

$$<F> = m\eta N V_0 \left[1 + \lim_{T\to\infty} \frac{1}{T}\int_{t_0}^{T} \frac{1}{N}\sum_{j=1}^{N}\left(\frac{\dot{X}_j}{V_0}-1\right)^2 dt\right] \quad (2).$$

Deriving this equation we took into account that the energy pumped in by the pulling force per unit time equals to the energy dissipated rate $m\eta \lim_{T\to\infty}\frac{1}{T}\int_{t_0}^{T}\sum_{j=1}^{N}\dot{X}_j^2\,dt$. For chains shorter than the critical length, $N_{cr}$, using a perturbation theory we can expand particles velocities, $\dot{X}_i$, in terms of the small parameter $\epsilon = K_{int}/K$, as $\dot{X}_i = \dot{X}_i^0 + \epsilon \dot{X}_i^1 + O(\epsilon^2)$, where $\dot{X}_0 = V_0$ is the exact solution for $\epsilon = 0$. Then, it is straightforward to find [22] that

$$<F> = Nm\eta V_0\left[1 + \alpha_1\left(\frac{K_{int}}{K}\right)^2 + O(\epsilon^3)\right] \quad (3).$$

While it remains difficult to get an analytical expression for $\alpha_1$, it is easy to extract its value from data fitting of the numerical simulations obtained for intermediate chain lengths $200 \lesssim N < N_{cr}$, where <F>/N is almost independent of N (Fig. 2(b)). It is evident from Fig. 4(a) that, except for one point corresponding to $\epsilon = 0.75$, the theoretical predictions for <F> agree well with the numerical results obtained for a wide range of values for $U_0$, $K$, $V_0$ and $\eta$. The discrepancy for $\epsilon = 0.75$ is understandable as in this case $\epsilon$ is no longer a small parameter.

While Eq.(3) describes well the average friction force for the chains shorter than the critical length, it does not work for longer chains which exhibit a transition to the commensurate state and the corresponding jump in the friction force. Our simulations show that above the critical length the friction force can be calculated as a sum of the contributions given by Eq.(3) and by the local region at the domain wall. These two contributions correspond to different modes of frictional motion, and depend differently on system parameters. Equation (3) describes "viscous-like" friction modified by the chain-substrate interactions that is proportional to $\eta$ and $V_0$ and only slightly depends on $K_{int}/K$. In contrast, the local contribution to $<F>$ results from the stick-slip motion, and it increases with $U_0$ and $K$ and practically independent of $\eta$ and $V_0$.

The results presented above show that the critical size of the slider exhibiting the super-low friction is limited by a nucleation of localized commensurate region that occurs for $N>N_{cr}$. In order to find the critical length above which the superlubricity disappears, we have to calculate the pulling force needed to induce the structural transition from the incommensurate to commensurate state. The applied force, $F_{ext}$, required to raise the distance between two rightmost particles, $i=N$ and $i=N-1$, from $a_c$ to $a_S$ can be estimated from the force balance at the $N$th particle that gives $F_{ext} = K(a_s - a_c) + \frac{U_0 2\pi}{a_S} \cos\left(\frac{2\pi x_N}{a_S}\right)$. Then the minimal value of this force reads as $F_{ext}^{cr} = K(a_s - a_c) - \frac{U_0 2\pi}{a_S}$. This equation, which does not include any fitting parameter, is in excellent agreement with the results of numerical simulations [19].

In the steady state regime of motion the time-averaged pulling force equals the friction force that depends on the length of the chain. Thus, the transition will occur when the friction force $<F>$ approaches $F_{ext}^{cr}$. Using the equation for $F_{ext}^{cr}$ and Eq.(3), we can calculate the critical length as

$$N_{cr} = L^{cr}/a_c = K(a_s - a_c)\left(1 - \beta \frac{K_{int}}{K}\right)/m\eta V_0\left[1 + \alpha_1\left(\frac{K_{int}}{K}\right)^2\right], \quad (4)$$

where $\beta = \frac{a_S}{2\pi(a_s - a_c)}$. Equation (4) shows that the critical length increases with the stiffness of inter-particle interaction within the chain and decreases with increasing the damping coefficient and pulling velocity. $N_{cr}$ also decreases with increasing the ratio, $K_{int}/K$, however this effect is weak since $K_{int}/K$ is a small parameter.

In Fig. 4(b), we compare the theoretical predictions of Eq.(4) with numerical results obtained for a broad range of values for $U_0$, $K$, $\eta$, and $V_0$. For all values of system parameters the theoretical results agree well with numerical simulations. We would like to stress that our model rely only on one fitting parameter, $\alpha_1 = 2.64 \pm 0.11$ that has been found from the comparison of analytical and numerical results for the average friction force.

We tested our theory by estimating the critical length for DWCNTs [11]. For this system the length, energy and time units used in the manuscript are $a_c$=0.2nm, $Ka_c^2$=49.7eV and $\tau_0 = \sqrt{m/K}$=80fs, respectively [19], and the range of dimensionless parameters studied in our numerical simulations corresponds to 0.1 eV <$U_0$ < 2.5 eV, 0.0375 ps$^{-1}$ < $\eta$ < 0.875 ps$^{-1}$ and 2.5 m/s < $V_0$ < 125 m/s. The analytical equation (4) can be used in much wider range of parameters that is inaccessible for simulations. We found that for DWCNTs $L_{cr}$ is about 50 times larger [19] than the experimentally investigated length range exhibiting superlubric regime [11], thus predicting the possibility for further scaling up of superlubricity.

Here, for clarity we considered the incommensurability of the chain and substrate structures corresponding to the golden ratio, for which the difference between two periods is relatively large. In this case the linear elastic description of the transition from the incommensurate to commensurate state considered in the Letter can break down. However, the results obtained are valid also for smaller misfits between the contacting lattices. Then the stretching of the chain at the transition is significantly smaller. It should be also noted that, our calculations of the critical length are based on the assumption that all particles of the chain experience the same microscopic friction proportional to the viscous damping coefficient. However, in realistic systems there could be additional contributions to the microscopic friction coming from chemical interactions between the edge particles and the surface and/or defects in the chain structure. These effects will lead to an increase of the average friction force thus reducing the critical length.

While we have not yet studied our model in either 2D or 3D systems, previous investigations of edge-driven extended elastic sliders [23-25] demonstrated that 1D consideration provides a good description of friction in 3D system. Thus, we expect that qualitative conclusions of this work will hold also for 3D systems, in particular for graphene layers deposited on substrates [26] and clusters studied in the context of nanomanipulation [27]. In the latter cases we expect shorter critical length-scale because of the larger contribution of high-friction boundaries compared to that in 1D systems.

**Acknowledgement**

We acknowledge Martin Müser for insightful comments and remarks. M. U. acknowledges support by the Israel Science Foundation Grant 1316 and by the German-Israeli Project Cooperation Program (DIP). A.B. and A.V. acknowledge support by the Swiss National Science Foundation Sinergia CRSII2 136287, and the ERC Advanced Grant No. 320796-MODPHYSFRICT. This work is also supported by COST Action MP1303.

**Captions**

**Fig. 1.** Schematic sketch of the experimental set-ups (a), and of the simulated model (b).

**Fig. 2.** (a) Instantaneous friction force normalized by the average single-particle viscous force for chains of 600 (black) and 1,000 particles (red). (b) Average friction force per particle normalized by the viscous force as a function of the chain length. The inset shows the total friction force as a function of the chain length. Parameter values used in simulations: $u_0 = 0.02$, $\gamma = 3.2 \times 10^{-2}$, $v_0 = 1.6 \times 10^{-2}$ and $k_{dr} = 0.1$.

**Fig. 3.** (a) The upper panel shows the friction force per particle versus time, the lower one shows the normalized distance between the two rightmost particles for a chain of 1,000 particles. (b) and (c) are maps of the normalized distances between particles as a function of time for $N = 600$ and $N = 1,000$ respectively. The inset to the panel 3(c) shows the distribution of distances between particles *vs* the particle index calculated in the steady state regime for the chains of $N=600$ (black) and 1,000 (red); (d) Normalized average dissipated power as a function of the scaled coordinate $x_{sc} = i/N$ for $N = 600$ (black) and $N = 1,000$ (red). Values of the rest parameters are the same as in Fig.2.

**Fig. 4.** (a) Comparison between the average friction force per particle $<F>/N$ calculated numerically and given by Eq.(3) with $\alpha_1 = 2.64 \pm 0.11$. Black squares, red circles, blue and purple triangle obtained by varying $0.02<U_0<0.5$, $5<K<100$, $0.005<V_0<0.1$ and $0.01<\eta<0.2$, respectively, and keeping other parameters from the set $U_0=0.2$, $K=10$, $V_0=0.05$ and $\eta=0.1$ constant. (b) Comparison between $L_{cr}$ calculated numerically and using Eq. (4).

**Figures**

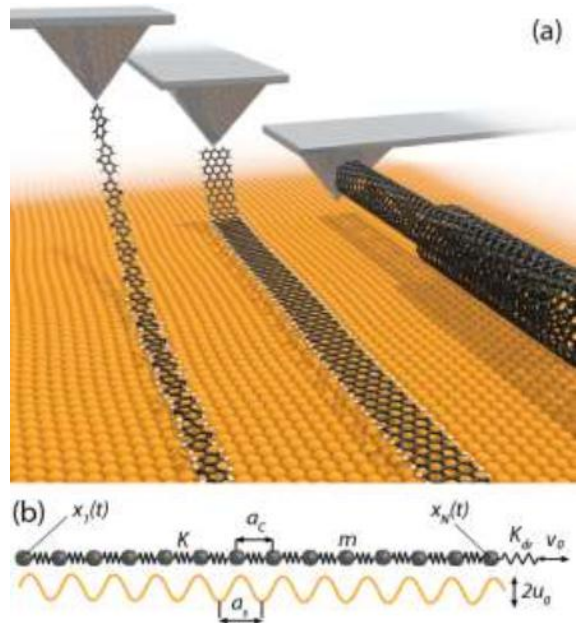

Fig. 1.

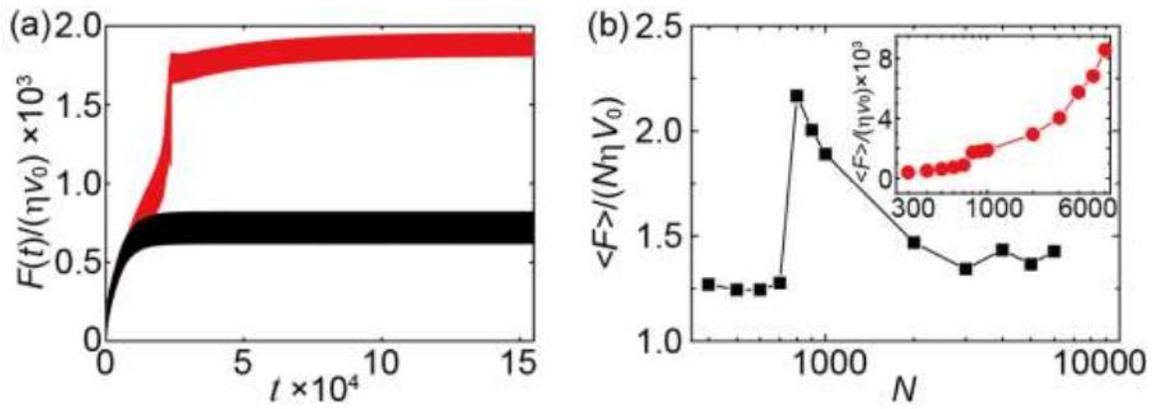

Fig. 2.

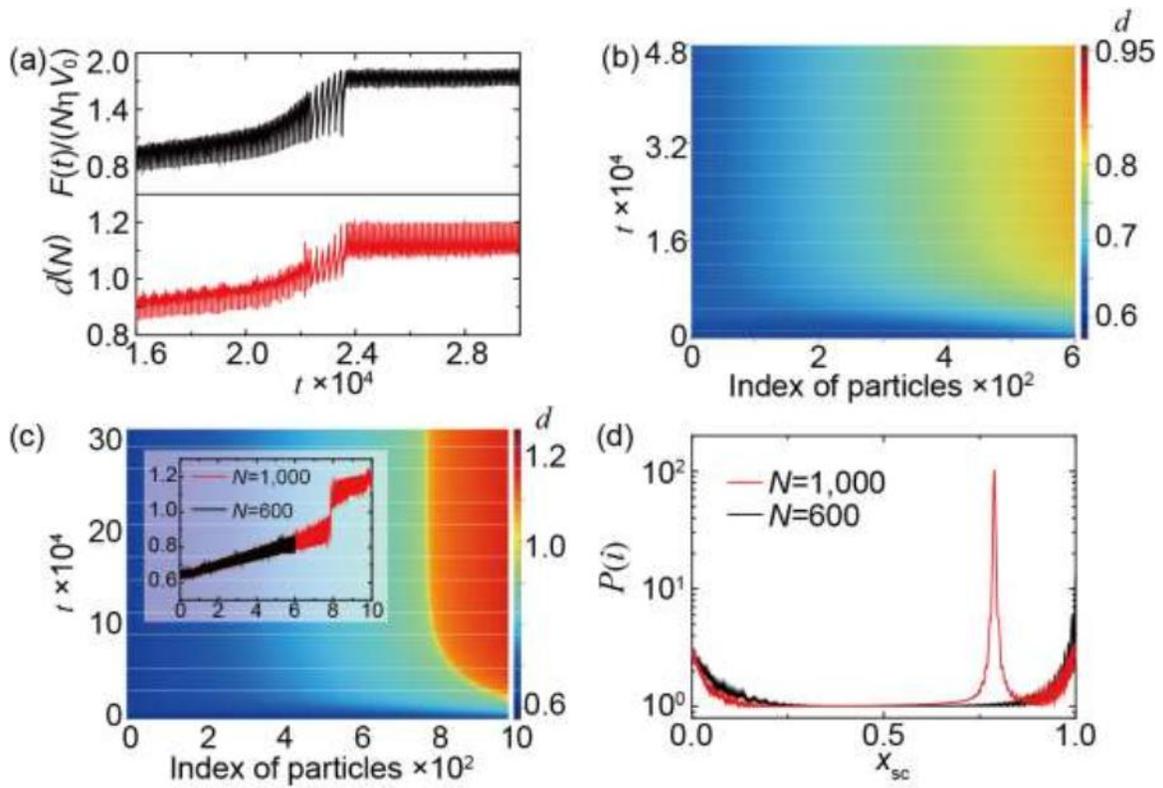

Fig. 3.

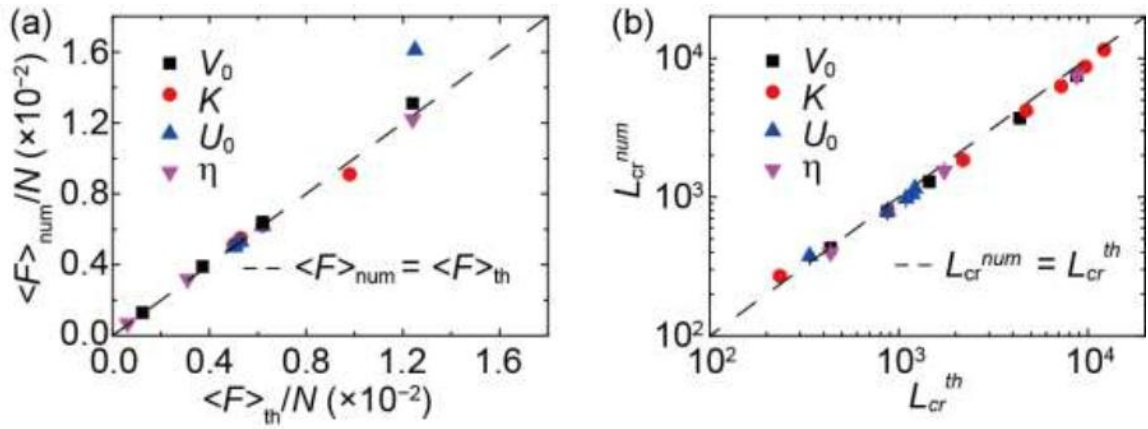

Fig. 4.

## Supplementary Material

This supplementary material contains the following parts

1. Check whether the theoretical model also applies to system where $a_c > a_s$.
2. Critical force for the transition to a high-friction state.
3. Details of the estimation of $L_{cr}$ for DWCNTs used experimentally.
4. Video materials.

### 1. Check whether the theoretical model also applies to system where $a_c > a_s$.

To check whether the theoretical model also applies to system where $a_c > a_s$, we conduct similar numerical calculations for $a_c/a_s = \frac{1+\sqrt{5}}{2}$. Sixteen sets of parameters are tested, where $U_0$ ranges from 0.025 to 0.1, $K$ ranges from 8 to 40, $\eta$ ranges from 0.015 to 0.15 and $V_0$ ranges from 0.015 to 0.1, and keeping other parameters from the set $U_0=0.076$, $K=10$, $\eta=0.1$ and $V_0=0.05$ constant. The average friction force is also fitted to $<F> = Nm\eta v_0 \left[1 + \alpha_1 \left(\frac{K_L}{K}\right)^2\right]$ based on the four points where $\epsilon = 0.1, 0.2, 0.3$ and $0.4$ only which results in $\alpha_1 = 2.69 \pm 0.15$. Then we plot this theoretical prediction for $<F>/N$ for the rest parameter sets and compare it with numerical results as shown in Fig. S1a. It is evident that the theoretical predictions agree well with numerical results. We then estimated the critical length which is normalized by $a_c$ using $K(2a_s - a_c)\left(1 - \beta\frac{K_L}{K}\right)/F_{avg}$ where $\beta = \frac{a_s}{2\pi(2a_s - a_c)}$ and compare it with numerical results as shown in Fig. S1b, which again shows good agreement. Thus, we conclude that this theoretical model also applies in systems where $a_c > a_s$.

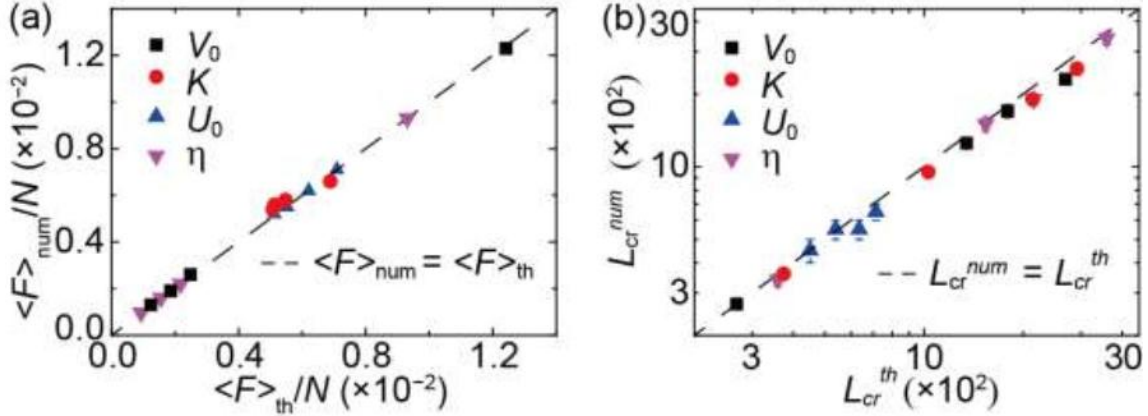

**Figure S1. Analytic expression for the critical length with $a_c > a_s$.** (a) Comparison between average friction force per particle $<F>/N$ calculated numerically and theoretically. Here $\alpha_1 = 2.69 \pm 0.15$ it fitted based on the four points $\epsilon = 0.1, 0.2, 0.3$ and 0.4 only. (b) Comparison between critical length $L_{cr}$ calculated numerically and theoretically, where no fitting parameter presents.

2. **Critical force for the transition to a high-friction state.**

The theory developed in the Letter predicts that the critical pulling force needed to induce the structural transition from the incommensurate to commensurate state can be calculated as

$$F_{ext}^{cr} = K(a_s - a_c) - \frac{U_0 2\pi}{a_S}$$

Here we demonstrate that this prediction agrees well with the results of numerical simulations. The figure R1 presented below show a comparison between the analytical equation for $F_{ext}^{cr}$, and numerical results for different values of the chain spring constant, $K$, and the amplitudes of the periodic substrate potential, $U_0$. The blue curves are numerical results for the instantaneous friction force $F(t)$, and the red horizontal lines are the predicted values of the critical force. We would like to remind that the sudden jump in the instantaneous friction force corresponds to the nucleation of a commensurate region, see Fig.3a in the Letter.

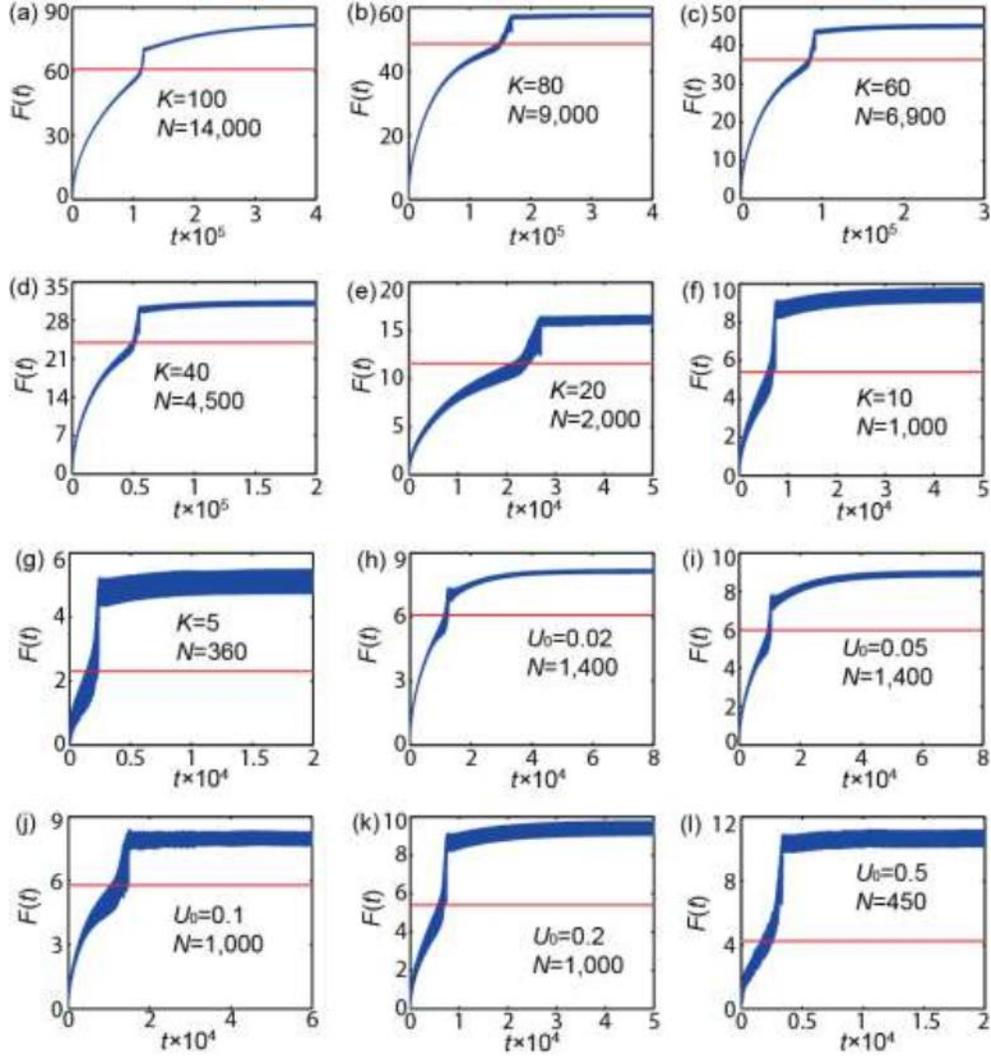

**Figure S2. Comparison between the analytical equation for $F_{ext}^{cr}$ and numerical results.** In panels, (a)-(g) $U_0$=0.2, and in panels (h)-(l) $K$=10, other parameters are as in the text.

One can see that for all values of parameters there is an excellent agreement between the theoretical predictions and simulations.

The theory also predicts that $F_{ext}^{cr}$ depends on the parameters $K$, $U_0$, $a_s$ and $a_c$, and is independent of $\eta$, $V_0$ and $N$. As shown in Fig. S3, this prediction is also in excellent agreement with the results of numerical simulations.

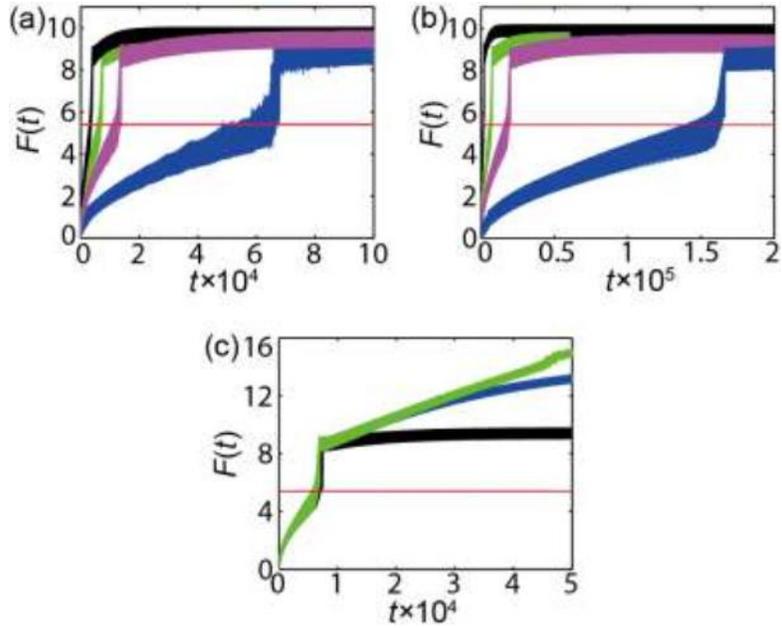

**Figure S3. The critical force is independent of $\eta$, $V_0$ and $N$.** Comparison between the results of simulations of $F(t)$ and the theoretical predictions for $F_{ext}^{cr}$ performed for different values of (a) $\eta$, (b) $V_0$ and (c) $N$. The curves show numerical results for $F(t)$ and the red horizontal lines are the theoretical values of $F_{ext}^{cr}$. (a) The black, green, magenta and blue curves are calculated for $\eta = 0.2$, $N = 540$; $\eta = 0.1$, $N = 1,000$; $\eta = 0.05$, $N = 2,000$ and $\eta = 0.01$, $N = 9000$, respectively. (b) The black, green, magenta and blue curves correspond to $V_0 = 0.1$, $N = 560$; $V_0 = 0.05$, $N = 1,000$; $V_0 = 0.03$, $N = 1,600$; and $V_0 = 0.01$, $N = 4,600$, respectively. (c) The black, blue and green curves correspond to $N = 1,000$; $2,000$ and $100,000$.

Thus, the equation for $F_{ext}^{cr}$ that does not include any fitting parameter captures the essence of the force induced transition from incommensurate to commensurate state.

## 3. Details of the estimation of $L_{cr}$ for DWCNTs used experimentally.

As an application of above consideration, we estimate the critical length for DWCNTs where the super-low friction has been observed up to centimeter length-scale [1]. For DWCNTs the intrashell and intershell stiffness can be estimated as $K=Ea_c$ and $K_s=Ga_s$ [2], respectively, where the Young's modulus $E$ ~0.5-1 TPa [3] and intershell shear modulus $G$ ~1 GPa [4] have been found experimentally. A value of 0.2 nm is used for $a_c$, which is a typical value for chemical bonds [2]. Experimental [1] and numerical [5] estimations of the damping coefficient give $\eta$ ~1 ps$^{-1}$. The mass, $m$, can be calculated from the bulk density of carbon nanotubes that gives $m = \rho_V \pi D h a_c$ where $\rho_V$=2,240 kg/m$^3$ is the bulk density, $h$=0.34 nm is the wall thickness [6] and $D$ is the diameter. For incommensurate DWCNTs, for instance for armchair tube confined in zigzag tube, $a_s-a_c$ is on the order of 0.1$a_{C-C}$, where $a_{C-C}$=0.142 nm is the carbon-carbon bond length. Thus, for DWCNTs with the diameter of the outer shell being ~2.73 nm pulled with a speed $V_0$ ~ 1 μm/s through a force probe attached to one its ends [1], $L_{cr}$ is estimated to be 0.5 m that is ~50 times larger than the length of DWCNT used in the experiments (~9 mm) [1]. Thus, our theoretical results agree with the experimental observations and predict the possibility for further scaling up of superlubricity.

### 4. Video materials.

The upper panel shows the edge of a 1,000 particles 1D chain driven through a spring attached to the rightmost atom. The $y$ coordinate of the particles is proportional to the surface potential they feel, this allows to visualize whether they sit in the minima, on the maxima or in between. The particles are colored according to their commensurability with respect to the substrate, more precisely according to the quantity $d(i)$ defined in the paper, red colors mean $d(i)$>1, blue colors mean $d(i)$<1. As the chain is driven a commensurate domain wall nucleates at the edge and propagates backwards. The second panel shows directly $d(i)$, with the gray dashed line representing a perfect commensurability. The third panel is the instantaneous energy dissipation and it easy to see how it localizes on the commensurate domain wall. The last panel represents the instantaneous friction force measured through the driving spring, the change in slope corresponds to the nucleation of the commensurate domain wall.